\documentclass[iop]{emulateapj}

\newcommand*{\teff}{$T_{\rm eff}$}
\newcommand*{\logg}{$\log~g$}

\newcommand*{\kms}{km s$^{-1}$}
\newcommand*{\zmax}{$Z_{\rm max}$}
\newcommand*{\rmax}{$r_{\rm max}$}
\newcommand*{\rmin}{$r_{\rm min}$}

\newcommand*{\vphi}{$V_{\rm \phi}$}
\newcommand*{\vtheta}{$V_{\rm \theta}$}

\newcommand*{\msun}{$M_\odot$}

\newcommand*{\rsun}{$R_\odot$}

\newcommand*{\stackel}{St$\ddot{a}$ckel}
\newcommand*{\gaia}{$Gaia$}
\newcommand*{\vrv}{$V_{\rm r}^{2}/V^{2}$} 

\usepackage{color}
\usepackage{epsfig}
\usepackage{graphicx}
\usepackage{natbib}
\usepackage{hyperref}

\hypersetup{
    bookmarks=true,         
    unicode=false,          
    pdftoolbar=true,        
    pdfmenubar=true,        
    pdffitwindow=true,     
    pdfstartview={FitH},    
    pdftitle={},    
    pdfauthor={Beers},     
    pdfsubject={Astronomy},   
    pdfcreator={dvipdf},   
    pdfproducer={dvipdf}, 
    pdfkeywords={metal-poor stars},
    pdfnewwindow=true,      
    colorlinks=true,       
    linkcolor=red,          
    citecolor=blue,        
    filecolor=magenta,      
    urlcolor=cyan,           
    breaklinks=true,
    linktocpage
}

\shorttitle{Multiple Accretion Events in GSE}
\shortauthors{Kim et al.}
\slugcomment{Draft version, March 20, 2021}

\begin{document}

\title{Evidence For Multiple Accretion Events in the
\gaia-Sausage/Enceladus Structures}

\author{Young Kwang Kim\altaffilmark{1}, Young Sun Lee\altaffilmark{1,3}, Timothy C. Beers\altaffilmark{2}, and Jae-Rim Koo\altaffilmark{1}}
\altaffiltext{1}{Department of Astronomy and Space Science, Chungnam National University, Daejeon 34134, Republic of Korea}
\altaffiltext{2}{Department of Physics and JINA Center for the Evolution of the Elements, University of 
                 Notre Dame, IN 46556, USA}
\altaffiltext{3}{Corresponding author; youngsun@cnu.ac.kr}

\begin{abstract}

We present evidence that multiple accretion events are required to explain
the origin of the \gaia-Sausage and Enceladus (GSE) structures,
based on an analysis of dynamical properties of main-sequence stars from the Sloan
Digital Sky Survey Data Release 12 and $Gaia$ Data Release 2. GSE
members are selected to have eccentricity ($e$) $>$ 0.7 and
[Fe/H] $<$ --1.0, and separated into low and high orbital-inclination
(LOI/HOI) groups. We find that the LOI stars mainly have $e < 0.9$ and
are clearly separable into two groups with prograde and retrograde
motions. The LOI stars exhibit prograde motions in the inner-halo region
and strong retrograde motions in the outer-halo region. We
interpret the LOI stars in these regions to be stars accreted from two 
massive dwarf galaxies with low-inclination prograde and
retrograde orbits, affected to different extents by dynamical friction
due to their different orbital directions. In contrast, the majority of
the HOI stars have $e > 0.9$, and exhibit a globally symmetric
distribution of rotational velocities (\vphi) near zero, although there
is evidence for a small retrograde motion for these stars (\vphi\
$\sim$ --15 $\rm{km~s^{-1}}$) in the outer-halo region. We consider these stars to
be stripped from a massive dwarf galaxy on a high-inclination
orbit. We also find that the LOI and HOI stars on highly eccentric and
tangential orbits with clear retrograde motions exhibit different
metallicity peaks at [Fe/H] = --1.7 and --1.9, respectively, and argue
that they are associated with two low-mass dwarf galaxies accreted in
the outer-halo region of the Galaxy.

\end{abstract}


\section{Introduction} \label{sec:intro}

The studies of the stellar halo of the Milky Way (MW), which is thought to be assembled
via multiple hierarchical mergers (\citealt{white1991}), provide
valuable clues to its formation and evolutionary history, as its long
dynamical timescale preserves a fossil record of past accretion events
(\citealt{Bland-Hawthorn2016}). The advent of the $Gaia$ Data Releases
($Gaia$ DRs; \citealt{gaia2016,gaia2018}), which provide precise
astrometric information for many millions of stars, has dramatically
expanded the detail of our view of the MW's accretion history. For
instance, the combination of $Gaia$ and large spectroscopic survey data
has enabled the detection of distinctive accretion signatures from the
$Gaia$-Sausage (GS; \citealt{belokurov2018}) and $Gaia$-Enceladus (GE;
\citealt{helmi2018}). Other small-scale accretion events have also been
discovered (\citealt{myeong2018,koppelman2019,myeong2019,naidu2020,necib2020,yuan2020,horta2021,refiorentin2021}).

According to the study of \citet{belokurov2018}, who used a sample of
main-sequence (MS) stars from the Sloan Digital Sky Survey (SDSS;
\citealt{york2000}) along with \gaia\ DR1 astrometry, the GS structure
exhibits strong radial anisotropy and a mildly prograde motion of $20
\sim 30$ \kms\ for stars with --1.7 $<$ [Fe/H] $<$ --1.0. Based on
cosmological zoom-in simulations of the formation of the stellar halo,
they argued that the GS is the result of an accretion event of a massive
dwarf galaxy with orbital eccentricity $e > 0.9$. From a sample of
nearby MS and blue horizontal-branch stars with SDSS DR9
(\citealt{ahn2012}) spectroscopy and $Gaia$ DR2 proper motions,
\citet{deason2018} demonstrated that the GS stars with --1.5 $<$ [Fe/H]
$<$ --1.0 have eccentricities $e > 0.9$ in the range of $10 <
r_{\rm{max}}$ (apogalactic distance) $< 30$ kpc, and an average
$r_{\rm{max}}$ well agrees with the break radius of the MW stellar
halo (\citealt{deason2013}).

From an analysis of the kinematics, chemical abundances, ages, and
spatial distributions of disk and halo stars in $Gaia$ DR2 with
available spectroscopy from the Apache Point Observatory Galactic
Evolution Experiment (APOGEE) DR14 (\citealt{majewski2017,
abolfathi2018}), \citet{helmi2018} identified that GE stars primarily
occupy the inner-halo region, and exhibit a small retrograde net motion.
They also suggested that the GE stars are debris from a massive dwarf
galaxy, inferred from their relatively low [${\alpha}$/Fe] and a large
spread in metallicity. \citet{mackereth2019} additionally found a robust
correlation between the chemical abundances and the orbital
eccentricities of local halo stars using the APOGEE DR14 data. Their
results indicated that the majority of local halo stars have highly
radial orbits with $e > 0.7$ (see also \citealt{mackereth2020}), and
relatively low abundances of [Mg/Fe], [Al/Fe], and [Ni/Fe], which are
comparable to those of stars observed in massive surviving
satellite galaxies of the MW.

The general consensus of the above studies is that both the GS and GE
stars exhibit eccentric and radial orbits in the inner halo, and they
are possibly the remnants of a single massive ($\sim 10^{9}$ \msun) disrupted
galaxy. Nonetheless, closer inspection has suggested that they have
different rotational motions, and that the GS stars appear to have $e > 0.9$,
higher than the GE stars (\citealt{belokurov2018, deason2018, helmi2018,
mackereth2019}). Thus, it is still an open question whether or not the
GS and GE stars have originated from a single accretion event.
 
Recent studies have demonstrated that the dynamical properties of likely
progenitors for different substructures can be a powerful tool to
distinguish one accretion episode from another. For example, studies of
minor-merging simulations (\citealt{read2008,villalobos2008,
jean-baptiste2017, karademir2019}) showed that the key ingredients of
merged galaxies to understand the orbits of stripped stars are their
orbital eccentricity, inclination, and inferred mass. In particular, the
orbital eccentricities of disrupted stars largely remain unchanged after
accretion (\citealt{mackereth2019}), and the majority of these stars
orbit with the same orbital inclinations as of their parent galaxies
(\citealt{refiorentin2015}).    

Retrograde motions of stars in the outer halo of a galaxy have been
shown by merging simulations to arise from dwarf parent galaxies on 
low-inclination retrograde orbits (\citealt{bignone2019}). Two massive
dwarf galaxies with different orbital directions on high-eccentricity
orbits (\citealt{murante2010}) can also produce stars with retrograde
motions. Observations also confirm these predictions by various simulations.
\citet{helmi2018} found, from isolated simulations of minor mergers
(e.g., \citealt{villalobos2009}), that the retrograde motion of the GE
stars is similar to that of a retrograde encounter with a
low-inclination dwarf. In addition, \citet{simion2019} verified that the
Hercules-Aquila cloud and Virgo Over-density are dominated by stars on
highly eccentric orbits, which are commensurate with the kinematic and
orbital properties of the GS stars. They also showed that both of the
diffuse debris clouds associated with this structure have high
orbital-inclination trajectories. These studies imply that dwarf
galaxies with different orbital inclination leave distinct dynamical
signatures in their disrupted stars; thus the orbital inclination of
their disrupted stars can be used to trace the orbital properties of
their progenitors.

Following the above reasoning, in this letter we first identify two
groups of stars --- low orbital-inclination (LOI) and high 
orbital-inclination (HOI) stars --- among stars with high eccentricity ($e >
0.7$) and [Fe/H] $<$ --1.0, the typical properties of the 
GS and GE (GSE, hereafter) stars, and report on the distinct 
chemical and dynamical signatures of
LOI and HOI groups, providing strong evidence for $multiple$
accretion events involved in the formation of the GSE structures.

This letter is organized as follows. We explain our sample selection in
Section~\ref{sec:sel}, and calculations of velocity components and
orbital parameters of our sample stars in Section~\ref{sec:orbit}. In
Section~\ref{sec:results}, we examine kinematic and orbital
properties of LOI and HOI stars in the inner-halo region, and present
evidence for accretion events that are distinct from the GSE
structures, as well as the identification of retrograde motions
associated with stars in the outer-halo region.
Section~\ref{sec:discussion} discusses the implications of our findings;
a summary follows in Section \ref{sec:summary}.

\section{Selection of Sample Stars} \label{sec:sel}

We have collected a sample of stars with available medium-resolution ($R
\sim 2000$) spectra from SDSS DR12 (\citealt{alam2015}), which includes
objects from the legacy SDSS program, the Sloan Extension for Galactic
Understanding and Exploration (SEGUE; \citealt{yanny2009}), and the
Baryon Oscillation Spectroscopic Survey (BOSS; \citealt{dawson2013}),
covering the extinction-corrected magnitude and dereddened color ranges
$14.0 < g_0 < 20.0$ and $0.0 < (g-r)_0 < 1.2$, respectively.

Using the SEGUE Stellar Parameter Pipeline (SSPP;
\citealt{allendeprieto2008, lee2008a,lee2008b}), we
estimated stellar atmospheric parameters (\teff, \logg, and [Fe/H])
for each star. In order to obtain reliable stellar-parameter estimates
for MS and MS turnoff (MSTO) stars, we restrict our analysis to stars
with average spectral signal-to-noise (S/N) ratios greater than 10.0,
$4400 \leq T_{\rm{eff}} \leq 7000$ K, and $\log~g \geq 3.5$. For stars
with multiple observations, we chose the star with the highest S/N, and
removed stars with apparently defective spectra. 
 
Proper motions for stars with errors less than 1.0 mas $\rm{yr}^{-1}$
were obtained through cross-matching with $Gaia$ DR2
(\citealt{gaia2018}). Radial velocities were adopted from the SDSS
pipeline; these have a precision of 5 -- 20 \kms, depending on the S/N
of the spectrum, and negligible zero-point errors (\citealt{yanny2009}).
For stellar-distance estimates, we employed the methods of
\citet{beers2000, beers2012}, as our program stars are mostly too faint
to have reliable parallaxes available from \gaia\ DR2. Their reported
uncertainty is on the order of 15 -- 20\%, as verified by comparing our
derived distances with $Gaia$ DR2 distances based on parallaxes with
relative errors less than 10\% (\citealt{kim2019, lee2019}) .

\section{Space Velocity Components and Orbital Parameters} \label{sec:orbit}

Given the distances, radial velocities, and proper motions adopted for
our sample of stars, we derived their space velocity
components in a spherical coordinate system. For these calculations, we
adopted $V_{\rm LSR}$ = 236 \kms\ (\citealt{kawata2019}) for the
rotation of the local standard of rest (LSR), a solar position of \rsun\
= 8.2 kpc (\citealt{Bland-Hawthorn2016}) in the disk plane from the
Galactic center, and a vertical distance of $Z_{\odot} = 20.8$ pc
(\citealt{bennet2019}) from the midplane. The solar peculiar motion with
respect to the LSR was assumed to be ($U$, $V$, $W$)$_{\odot}$ =
(--11.10, 12.24, 7.25) \kms\ (\citealt{schonrich2010}), where the
velocity components $U$, $V$, and $W$ are positive in the direction
toward the Galactic anticenter, Galactic rotation, and north Galactic
pole, respectively. In our adopted system, a star with \vphi\ $ > 0$
\kms\ has a prograde motion; retrograde rotation is indicated by
\vphi\ $ < 0$ \kms. Stars with $V_{\rm r} > 0$ \kms\ move away from the
Galactic center, and stars with \vtheta\ $> 0$ \kms\ move toward the
south Galactic pole.

\begin{figure}[t]
\epsscale{0.6}
\plotone{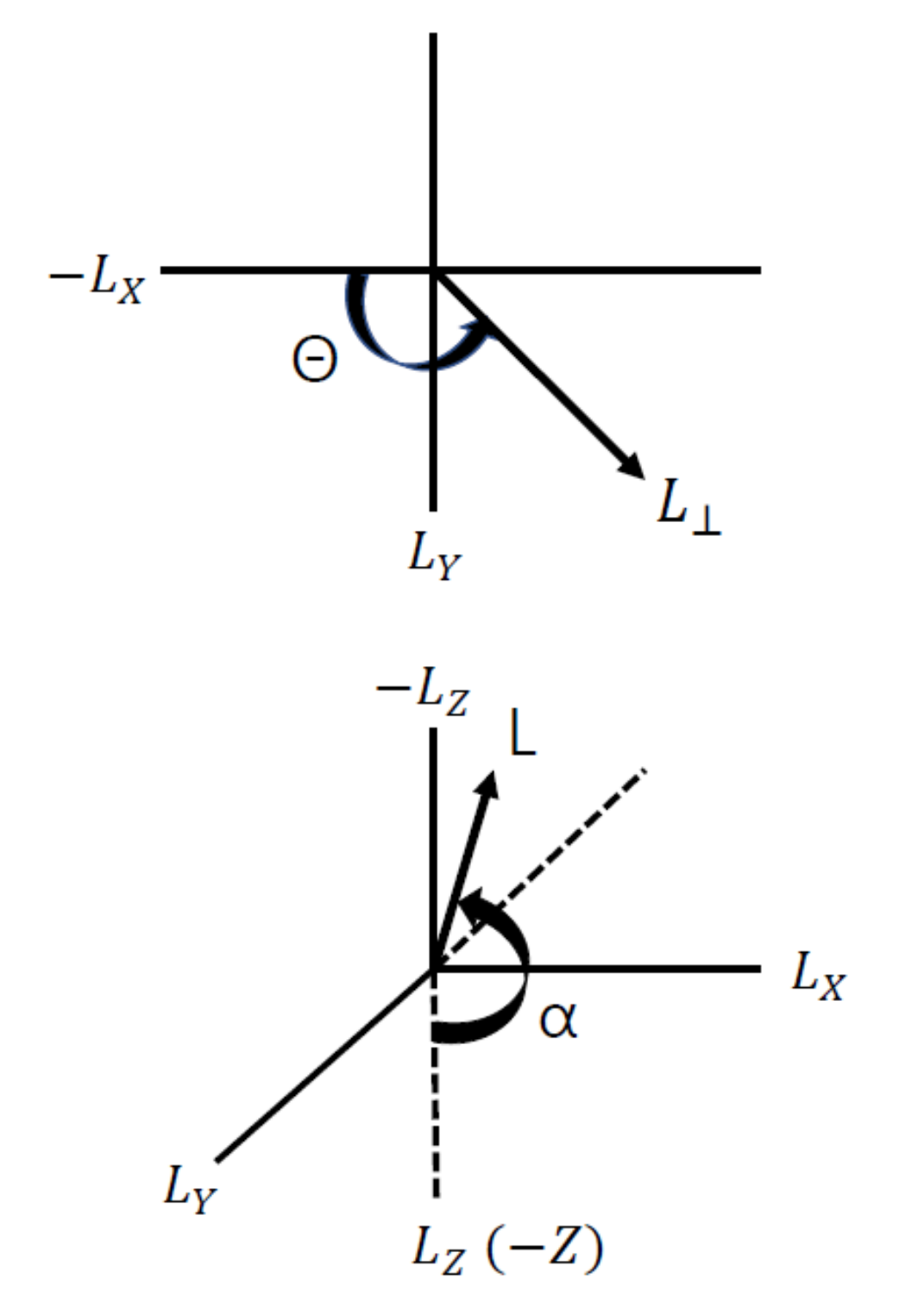}
\caption{Top: Position angle (${\Theta}$) of $L_{\bot}$, measured 
from the negative $L_X$ axis to the positive $L_Y$-axis direction. 
Bottom: Angle ($\alpha$) between $L$ and $L_Z$, measured from the positive $L_Z$ axis}
\label{figure1}
\end{figure}

We also made use of a Galactocentric Cartesian reference frame, denoted
by $(X,Y,Z)$, where the axes are positive in orientation toward the Sun,
Galactic rotation, and north Galactic pole, respectively. In addition,
we introduced an angle ($\alpha$) between the orientation of the total angular
momentum vector and the $L_Z$ axis, and a position angle
(${\Theta}$) of $L_{\bot}$ measured from the negative $L_X$ axis to the
positive $L_Y$-axis direction of the angular momentum component in the
$X$-$Y$ plane. These are defined by:

\begin{eqnarray*}
{\alpha} = {\cos}^{-1} \Big( \frac{L_Z}{L} \Big) ~~{\rm{and}}~~ {\Theta} = {\tan}^{-1} \Big(- \frac{L_Y}{L_X} \Big),
\end{eqnarray*}

\noindent where $L = \sqrt{L_{\bot}^2 + L_Z^2}$ is the total angular momentum, 
$L_{\bot} = \sqrt{L_X^2 + L_Y^2}$; $L_X$, $L_Y$, and $L_{Z}$ are the
$X$, $Y$, and $Z$ components of the angular momentum, and are positive
along the positive $X$-axis, the positive $Y$-axis, and the negative
$Z$-axis directions, respectively. Schematic representations 
for $\alpha$ and $\Theta$ are shown in Figure~\ref{figure1}. In this notation, stars with $L_{Z} >
0$ have prograde orbits and $\alpha < 90^{\circ}$. Retrograde orbits
have $\alpha > 90^{\circ}$, and the inclination angle ($i$) of their
orbital plane increases as $\mathrm{{\alpha}}$ approaches $90^{\circ}$.
For prograde motions, the inclination angle $i$ = $\alpha$, whereas $i$
= $180^{\circ}-{\alpha}$ for retrograde orbits.

\begin{figure}[t]
\epsscale{1.0}
\plotone{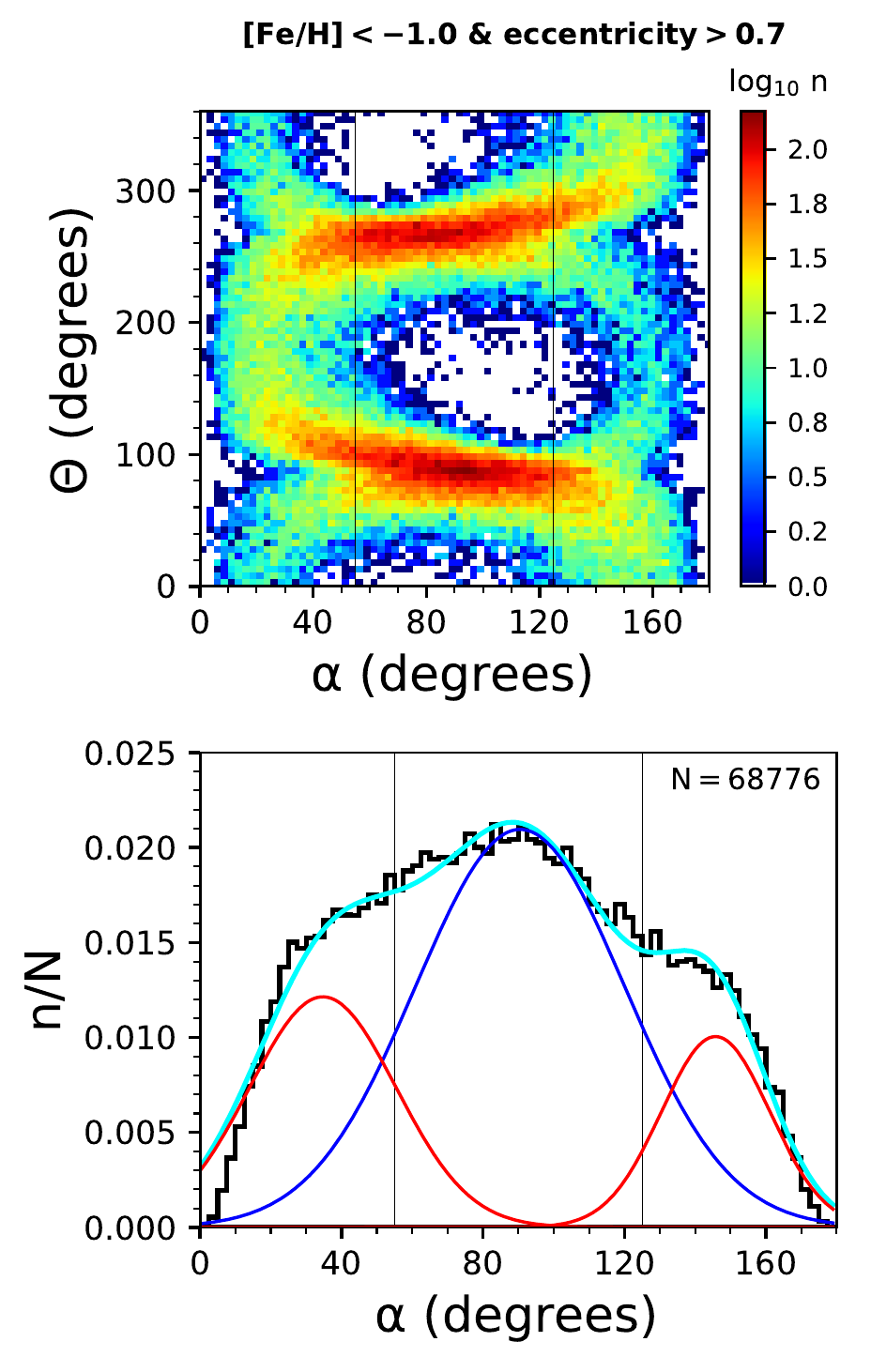}
\caption{Top panel: Map of the logarithmic number 
density in $\Theta$ versus $\alpha$ for stars with [Fe/H] $<$ --1.0 
and $e$ $>$ 0.7. Bottom panel: Histogram of $\alpha$ for our 
program stars (black line). The cyan solid line, which well-matches 
the black histogram, represents the sum of one normal 
distribution (blue line) and two (red). The vertical lines in 
both panels are marked at $\alpha = 55^\circ$ and $\alpha = 125^\circ$, 
respectively.}
\label{figure2}
\end{figure}

We employed an analytic \stackel-type potential (see \citealt{chiba2000,
kim2019} for details) in order to calculate the orbital parameters of
our sample stars, including the perigalactic distance (\rmin, the
minimum distance of an orbit from the Galactic center), apogalactic
distance (\rmax, the maximum distance of an orbit from the Galactic
center), and stellar orbital eccentricity ($e = (r_{\rm{max}} -
r_{\rm{min}}) /(r_{\rm{max}}+r_{\rm{min}})$), as well as \zmax\ (the
maximum distance of a stellar orbit above or below the Galactic plane).

Uncertainties on the derived kinematic and orbital values are obtained
from 100 Monte Carlo simulations with adopted uncertainties of 20\% in
distance, and quoted uncertainties in the radial velocity and proper
motions, assuming Gaussian error distributions. Prior to examining the
metallicity distribution function (MDF) of our program stars, we derived
simple selection functions to correct for the target-selection bias. The
selection function is defined by the fraction of the spectroscopically
targeted stars among the photometrically available targets in bin of 0.2
and 0.05 mag for a color-magnitude diagram of $r_{\rm 0}$ and $g_{\rm
0}-r_{\rm 0}$, separately for each SDSS plug-plate, as described in
\citet{lee2019}. 

Even though we obtained a total number of $N = 328,102$ stars with valid
orbital parameters, in this study we focus on the lower metallicity
stars with high-eccentricity orbits, in the range of [Fe/H] $<$ --1.0 and
$e > 0.7$, in order to reduce contamination from disk stars heated by 
the GSE (\citealt{belokurov2020}), and include the GSE stars according to \citet{mackereth2019}, 
resulting in a total of $68,776$ stars to be explored. Note that 
all of our selected stars are also included in the GSE even using 
the selection criteria by \citet{naidu2020}.

\begin{figure}[t]
\epsscale{1.0}
\plotone{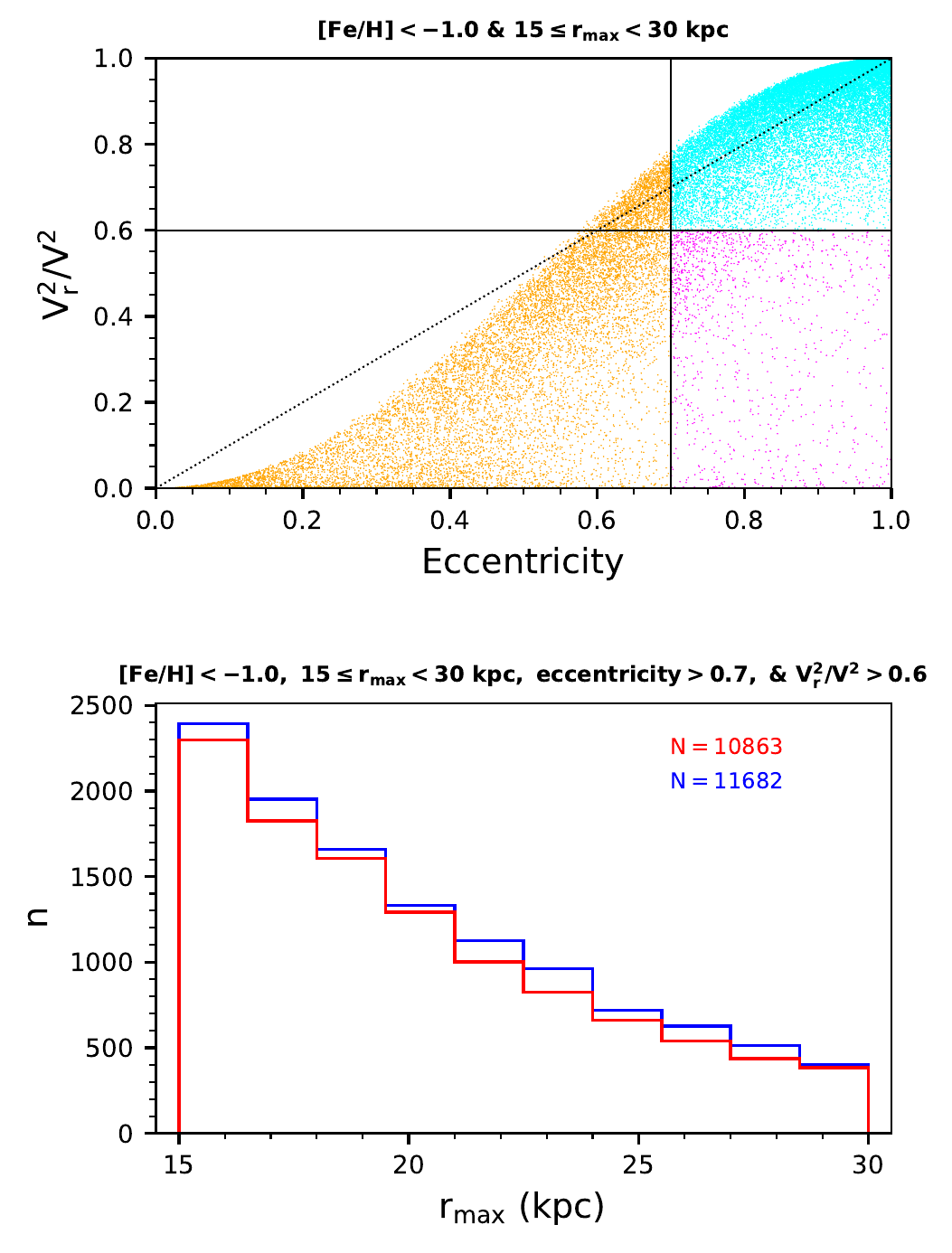}
\caption{Top panel: Distribution of $V_{\rm r}^2/V^2$ versus orbital eccentricity for stars 
with [Fe/H] $<$ --1.0 and $15 \leq r_{\rm{max}} < 30$ kpc. Cyan and magenta
dots represent stars on radial and tangential orbits, respectively, with
$e > 0.7$, while orange dots indicate stars with $e \leq 0.7$. 
Bottom panel: Histograms of \rmax\ for LOI (red) and HOI (blue) stars.}
\label{figure3}
\end{figure}

\section{Results} \label{sec:results}

In this section, we analyze the kinematic and orbital properties of
stars with highly eccentric and radial orbits in two regions: the
inner-halo region (IHR; $15 \leq r_{\rm{max}} < 30$ kpc) and the
outer-halo region (OHR; \rmax\ $\geq$ 30 kpc). In the IHR, we search for any 
differences in the dynamical properties between the LOI and HOI stars, while 
in the OHR, we explore the mean rotational velocity as a function of mean \rmax, and
the stellar MDFs for the LOI and HOI populations.

\subsection{Definition of LOI and HOI Populations} \label{subsec:def}

The top panel of Figure \ref{figure2} presents the logarithmic 
number density of the selected program stars in $\Theta$ versus $\alpha$ plane. In this 
map, we clearly see distinct kinematic features. Stars with relatively 
HOI of $55^\circ < \alpha < 125^\circ$ are mostly concentrated 
around ${\Theta} = 90^{\circ}$ and $270^{\circ}$, and have the highest 
density at ${\alpha} = 90^{\circ}$, whereas stars with relatively LOI 
of $\alpha \leq 55^\circ$ or $\alpha \geq 125^\circ$ are distributed 
over all ranges of ${\Theta}$. Moreover, the $\alpha$ histogram 
(black color) for our program stars, shown in the bottom panel
of Figure~\ref{figure2}, is well-fit with the sum (cyan line) of three different 
Gaussian distributions, which represent one (blue) for HOI and two (red) for LOI 
population. By assembling these characteristics together, we define LOI 
stars as those in the range $\alpha \leq 55^\circ$ or $\alpha \geq 125^\circ$; 
HOI stars are those in the range $55^\circ < \alpha < 125^\circ$. The 
condition of $55^\circ < \alpha < 125^\circ$ is equivalent 
to $L_{\bot}^2/L^2 > \sin^2 55^\circ = 0.67$, or 
$L_{\bot} > \pm \tan55^\circ L_Z = \pm 1.43 L_Z$ in the $L_Z$ versus $L_{\bot}$ space.

\subsection{Separation of GSE Stars into Low and High Orbital-Inclination Groups} \label{subsec:Sep}

First, in order to select GSE stars having highly eccentric and radial
orbits in the IHR, we plot stars in the ranges of $15 \leq r_{\rm{max}}
< 30$ kpc (\citealt{deason2018}) and [Fe/H] $<$ --1.0 in the plane of
\vrv\ versus orbital eccentricity, as shown in the top panel of
Figure~\ref{figure3}. In this figure, cyan, magenta, and orange dots
represent stars with $e > 0.7$ on radial and tangential orbits, and
stars with $e \leq 0.7$, respectively. Then, we choose GSE members (cyan
dots in the panel), defined by $e > 0.7$ (\citealt{mackereth2019}) and
\vrv\ $>$ 0.6 (\citealt{belokurov2018}), where \vrv\ $>$ 0.6 is equal to
$\beta > 0.67$. The velocity anisotropy parameter ($\beta$) of each star
is defined as $\beta$ = 1 -- $V_{\rm t}^2/2V_{\rm r}^2$ (see
\citealt{binney2008,elias2020}). 

Next, we divide the GSE members into two groups of stars: the
LOI and HOI sub-samples, by following the definitions 
described in Section \ref{subsec:def}. The bottom panel of Figure~\ref{figure3}
shows the \rmax\ distribution of the LOI (red) and HOI (blue)
populations, respectively. As can be seen, there are nearly equal
numbers of stars for the LOI and HOI populations as a function of \rmax. 
This indicates that their progenitor galaxies are likely different, 
as it is unlikely to have equal numbers of LOI and HOI stars 
if they are accreted from a single dwarf galaxy.

\begin{figure}[t]
\epsscale{1.2}
\plotone{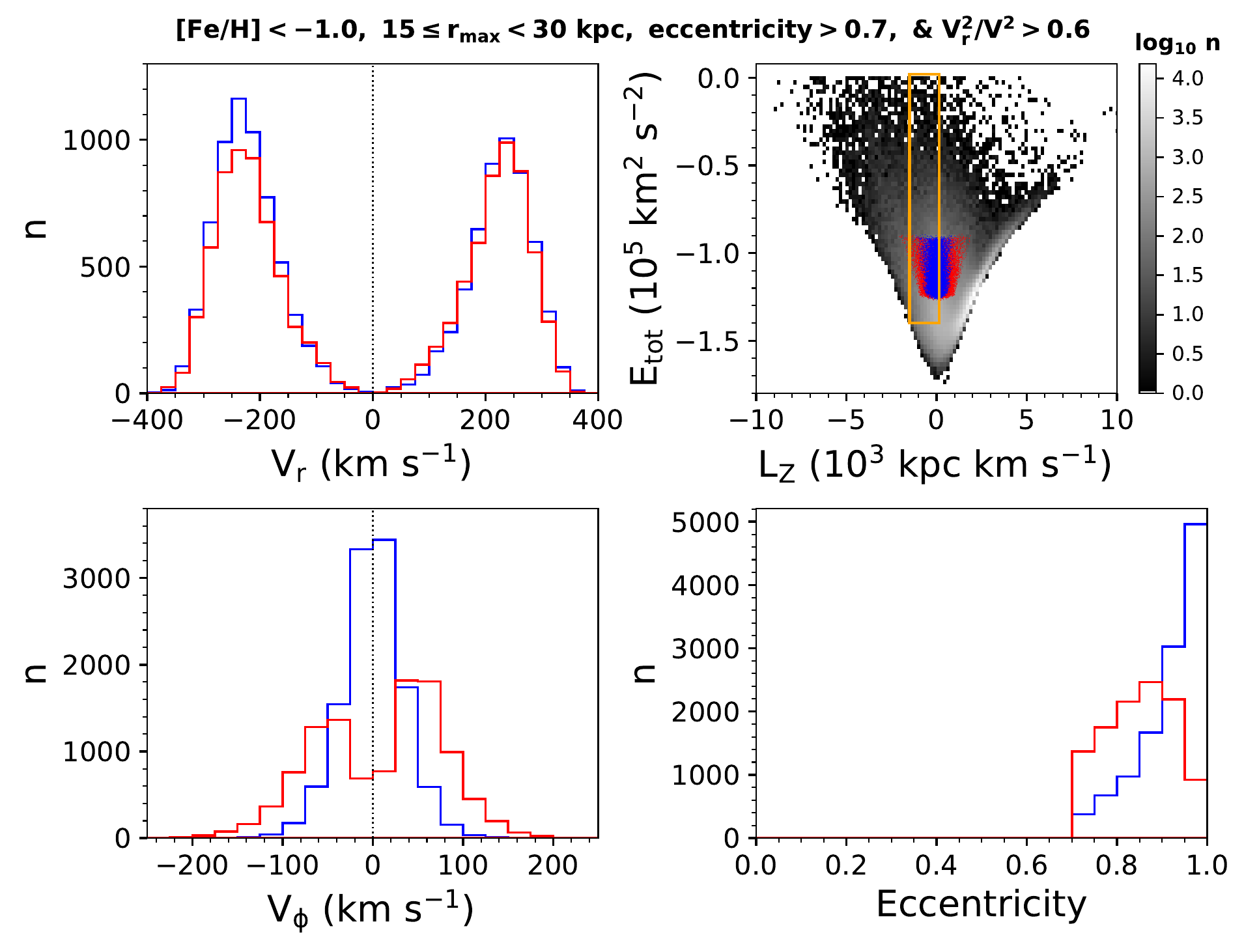}
\caption{Left column: Velocity distributions of $V_{\rm r}$ and $V_{\rm \phi}$ for LOI (red) 
and HOI (blue) stars, in the top and bottom rows, respectively. Right column: Same as 
in the left column, but for distributions of total energy ($E_{\rm tot}$) 
versus $L_Z$ (number densities are shown on a log$_{10}$ scale with gray 
color as indicated in the bar), and orbital eccentricity, in the top and bottom rows, respectively.
An orange solid box in the top-right panel marks the location of 
GE by \citet{helmi2018}, scaled to the Galactic potential used by this study.} 
\label{figure4}
\end{figure}

\subsection{Inner-Halo Region} \label{subsec:IHR}

The orange solid box in the top-right panel of Figure \ref{figure4} 
indicates the GE stars identified in \citet{helmi2018}. We note in the panel 
that the GE stars in the box are mostly made up of LOI and HOI stars 
with retrograde motions. Furthermore, we can observe that the HOI stars (blue histogram) 
have mostly $e > 0.9$ in the bottom-right panel of Figure~\ref{figure4}, which 
is characteristic of the GS stars reported by \citet{deason2018}. 
Thus, we can confirm that most of our selected LOI (red dots) and HOI (blue dots) 
stars follow typical properties of the GSE stars in $E_{\rm tot}$ versus $L_Z$ plane, 
although we applied slightly different selection criteria for the GSE member stars 
from the original works (e.g., \citealt{helmi2018}). As a result, our selection 
criteria for the GSE members do not alter the interpretation of the dynamical 
properties of the genuine GSE stars.

The top-left panel of Figure \ref{figure4} shows that the LOI and HOI sub-samples have essentially
identical distributions of $V_{\rm r}$. However, the \vphi\ distribution
(bottom-left panel) and the $e$-distribution (bottom-right panel) tell
an entirely different story -- the eccentricity distributions exhibit
completely different behaviors between the LOI (red) and HOI (blue)
populations. The LOI stars (red) are well-separated into two
sub-groups of prograde and retrograde motion in the \vphi\ distribution,
and the LOI group mainly occupies the range of $e <$ 0.9, while the
majority of stars in the HOI population have $e >$ 0.9. These properties
provide clear evidence that these stars have experienced different
accretion episodes. The HOI stars (blue) have almost zero net
rotational velocity, suggesting that they share a common progenitor.

\begin{figure}
\begin{center}
\epsscale{1.0}
\plotone{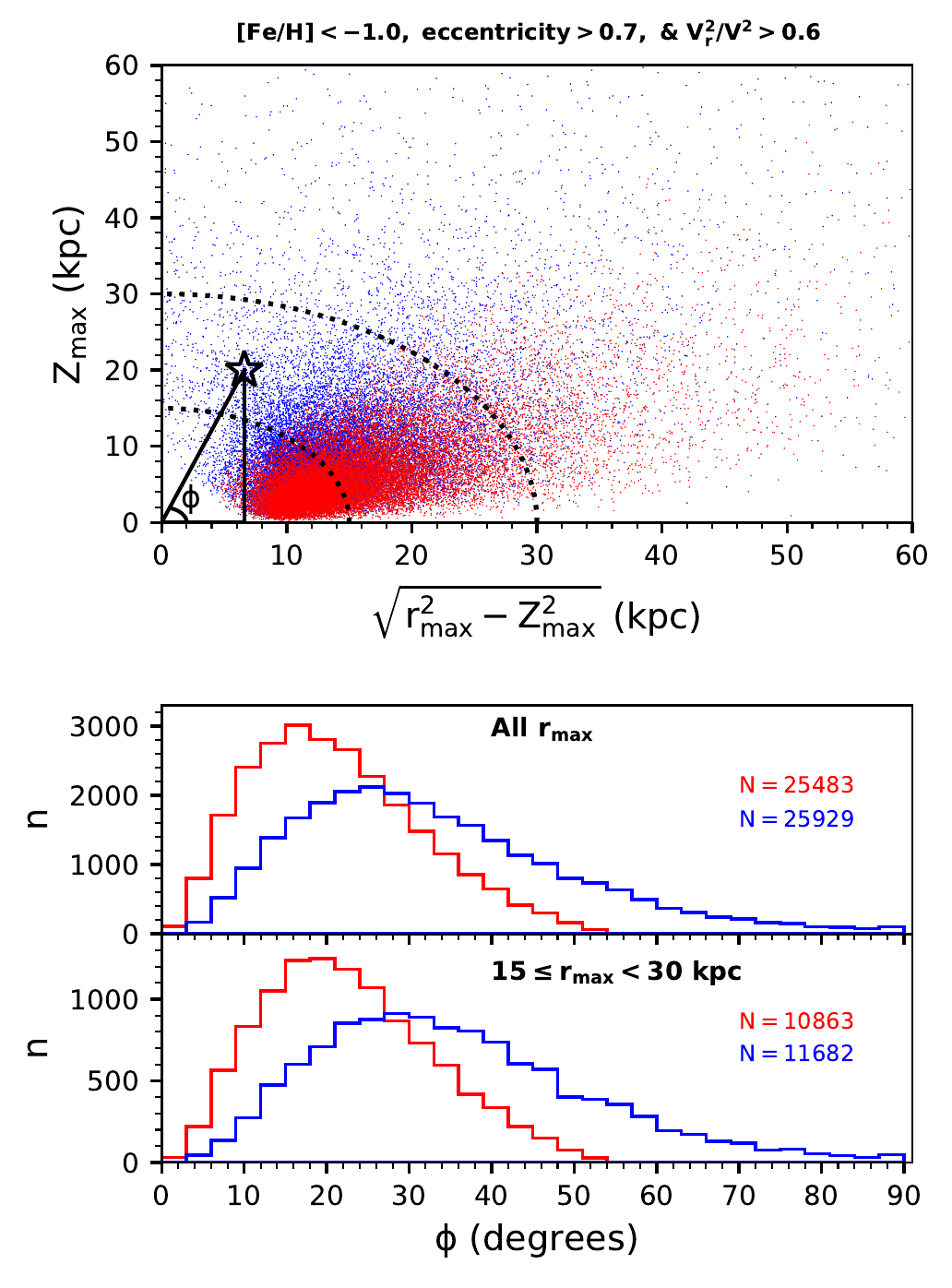}
\caption{Top panel: Distribution in \zmax\ versus $\sqrt{r_{\rm{max}}^2 - Z_{\rm{max}}^2}$ 
for LOI (red dots) and HOI (blue dots) stars on highly eccentric and radial orbits among 
our program stars. Inner and outer dotted curves mark with \rmax\ = 15 and 30 kpc, respectively. 
The right triangle is the schematic diagram for the definition of an angle,
$\phi$ = ${\tan}^{-1}(Z_{\rm{max}}/\sqrt{r_{\rm{max}}^2 - Z_{\rm{max}}^2})$. 
Middle panel: Distributions of $\phi$ for LOI (red) and HOI (blue) stars over 
the full range of \rmax. Bottom panel: Same as in the middle panel, 
but for $15 \leq r_{\rm{max}} < 30$ kpc.}
\label{figure5}
\end{center}
\end{figure}

We can use other orbital parameters, e.g., \zmax\ and
\rmax, to explore more the discrete dynamical signatures of
the LOI and HOI populations. To accomplish this, we first introduce 
an angle, $\phi$ = ${\tan}^{-1}(Z_{\rm{max}}/\sqrt{r_{\rm{max}}^2 - Z_{\rm{max}}^2})$.
As illustrated in the top panel of Figure \ref{figure5}, which shows 
the distribution in \zmax\ versus $\sqrt{r_{\rm{max}}^2 - Z_{\rm{max}}^2}$ for 
LOI (red dots) and HOI (blue dots) stars on highly eccentric and radial orbits, 
this angle is measured from the axis of $\sqrt{r_{\rm{max}}^2 - Z_{\rm{max}}^2}$
to a line connecting the origin to a star in the coordinate plane of
$\sqrt{r_{\rm{max}}^2 - Z_{\rm{max}}^2}$ and \zmax. Generally, this 
angle ($\phi$) increases with increasing \zmax\ at a 
given \rmax\ (or $E_{\rm{tot}}$). 

The middle and bottom panels of Figure \ref{figure5} 
show the distributions of the $\phi$ angles for LOI (red) and HOI (blue) stars, in 
the two regions of \rmax: all \rmax\ (middle panel) and $15 \leq r_{\rm{max}} < 30$ 
kpc (bottom panel). These panels immediately indicate that the HOI stars have mostly 
higher $\phi$ values than those of the LOI stars, both over all \rmax\ and $15
\leq r_{\rm{max}} < 30$ kpc ranges. Once again, this discrepancy
suggests a dynamical distinction between the two populations. One can
assume that if both HOI and LOI stars were accreted from a relatively
massive dwarf galaxy, as claimed in the literature (e.g.,
\citealt{belokurov2018, helmi2018}), no matter what the orbital
inclination they have, both groups of stars should exhibit similar
$\phi$ distributions, regardless of their \rmax. 

\begin{figure*}[!t]
\epsscale{1.15}
\plotone{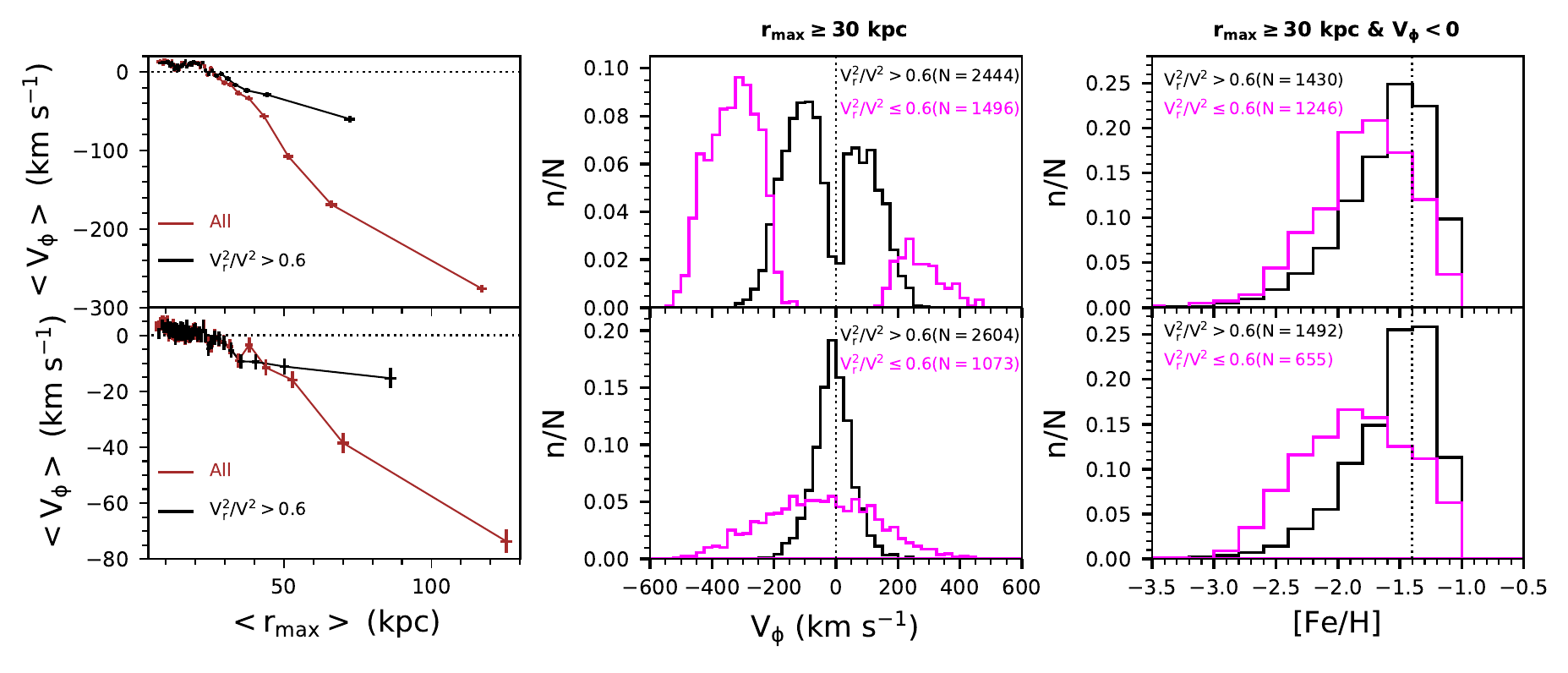}
\caption{Top panels: From left to right, profiles of mean rotational 
velocity, as a function of mean \rmax, for the full sample (brown color)
and for stars having $V_{\rm r}^2/V^2 > 0.6$ (black), distributions of rotational velocity for 
stars with \rmax\ $\geq$ 30 kpc, and MDFs for stars in ranges of \rmax\ $\geq$ 30 kpc 
and $V_{\phi} < 0$ for LOI stars with [Fe/H] $<$ --1.0 and $e$ $>$ 0.7. Bottom panels: 
Same as in top panels, but for HOI stars. In both panels, each mean value is 
obtained by passing a box of 500 stars in \rmax. The error bars on \vphi\ 
are obtained from 100 Monte Carlo samples. The black and magenta histograms in the \vphi\ 
distributions and MDFs represent stars separated into $V_{\rm r}^2/V^2 > 0.6$ 
and $V_{\rm r}^2/V^2 \leq 0.6$, respectively.}
\label{figure6}
\end{figure*}

\subsection{Outer-Halo Region} \label{subsec:OHR}

We now extend our search for diverse dynamical signatures between LOI
and HOI stars to the OHR, the region with \rmax\ $\geq$ 30 kpc. Once
again, we only consider stars with [Fe/H] $<$ --1.0 on highly eccentric
($e > 0.7$) orbits, and apply the selection criteria 
for LOI and HOI stars described in Section \ref{subsec:def} to the OHR. First, 
we search for evidence of retrograde motions and dissimilar MDFs for stars in the OHR.
The left plot of top panels in Figure~\ref{figure6} shows the profiles of
mean rotation velocity, as a function of \rmax, for the full sample of
stars (brown), and for stars with $V_{\rm r}^2/V^2 > 0.6$ (black). The
middle plot shows the \vphi\ distributions for stars with \rmax\ $\geq$
30 kpc. The right plot represents the MDFs for the stars with \rmax\
$\geq$ 30 kpc and \vphi\ $<$ 0 \kms. The black and magenta histogram in
the middle and right plots in each panel denote stars with $V_{\rm
r}^2/V^2 > 0.6$ (radial-motion dominated) and $V_{\rm r}^2/V^2 \leq 0.6$
(tangential-motion dominated), respectively. Plots in top panels are for
the LOI population, whereas bottom panels are same as in the top panels, but for
the HOI population.  

Inspection of the left plot of top panels shows that the LOI stars with
strong radial motions (black) exhibit stronger retrograde motions
in the OHR than in the IHR, and drops down to \vphi\ $\sim$ --70 \kms\
at \rmax\ $\sim 75$ kpc. This behavior can be also inferred from the
\vphi\ distribution (black histogram) in the middle plot, which presents
more stars with retrograde motion than in prograde motion. By way of contrast, the
HOI group of stars with strong radial motion (black) in the left
plot of bottom panels exhibits a small retrograde motion, \vphi\ $\sim$ --15
\kms, in the OHR, and does not change with increasing \rmax\ up to $\sim
85$ kpc. The middle plot indicates that their \vphi\ distribution (black
histogram) appears to be symmetric around \vphi\ $\sim 0$ \kms.
As a result, we realize that the behaviors in the \vphi\ distributions 
for both LOI and HOI stars with $V_{\rm r}^2/V^2 > 0.6$ in the OHR are 
not much different in the IHR. One more interesting aspect is that we 
generally see a steeper gradient (brown) of \vphi\ over \rmax\ in the OHR 
in the left plot of top and bottom panels. This is driven by the stars on 
the tangential orbits ($V_{\rm r}^2/V^2 \leq 0.6$) with retrograde motion.

We now examine the MDF of stars with retrograde motions in the OHR.
From the right plot of top and bottom panels of Figure~\ref{figure6}, one
can clearly see different MDFs between the radial- and tangential-dominated
samples for the LOI and HOI stars. The peaks of the MDFs for LOI stars
on highly radial and tangential orbits are at [Fe/H] = --1.5 and --1.7,
respectively, while the peaks of the HOI stars are at [Fe/H] = --1.3 and
--1.9, respectively. Converting to the stellar masses of 
their progenitors by the mass-metallicity relationship 
for dwarf galaxies (\citealt{kirby2013}), we obtained $M_{*} \sim 2 \times 10^{6}~M_{\odot}$ 
and $M_{*} \sim 4 \times 10^{5}~M_{\odot}$ 
for LOI stars, and $M_{*} \sim 3 \times 10^{6}~M_{\odot}$ and 
$M_{*} \sim 3 \times 10^{5}~M_{\odot}$ for HOI stars, respectively. 
Similarly, the stellar mass for the progenitor of the LOI stars 
with prograde motion on highly radial orbits is $M_{*} \sim 3 \times 10^{6}~M_{\odot}$.
Consequently, the discrepancies in the MDFs and progenitor 
masses suggest that the progenitor of the stars on highly radial 
orbits experienced different star-formation histories from that of those 
on tangential orbits, for both the LOI and HOI sub-samples.

\section{Discussion} \label{sec:discussion}

In the previous section, we showed that the LOI and HOI stars in the IHR
exhibit different distributions in \vphi, $e$, and the $\phi$ angle.
The LOI stars include more objects with $e < 0.9$ and small $\phi$
angles, and they are well-separated into stars with prograde and
retrograde motions. On the other hand, the HOI stars have more objects
with $e > 0.9$ and high $\phi$ angles, and they exhibit a single 
\vphi-distribution with a peak at \vphi\ $\sim$ 0 \kms. 

For stars in the OHR, the objects with tangential orbits among the LOI
population exhibit stronger retrograde motions than the HOI population.
Considering the stars with retrograde motions in the OHR, the
tangential-dominated sample exhibits a more metal-poor distribution than
the radial-dominated one for both the LOI and HOI populations.

How do we interpret our findings in terms of the accretion history of
the Galactic halo? Let us first consider the dynamical signatures of
merging galaxies predicted from various numerical simulations. It is
well-known that tidally stripped stars follow the orbital trajectories
of their parent galaxies when they are fully accreted into the MW (e.g.,
\citealt{quinn1986,vandenbosch1999,amorisco2017}) -- the accreted
stars preserve the orbital eccentricity (\citealt{mackereth2019}) and
inclination (\citealt{refiorentin2015}) of their parent galaxies.
Numerical simulations also predict stars with retrograde motions in the
outer-halo region of a galaxy, accreted not only from the merging of two 
massive dwarf galaxies with different orbital directions on
highly eccentric and low-inclination orbits (\citealt{murante2010}),
owing to the dissimilar efficacy of dynamical friction
(\citealt{quinn1986}), but also from one merged dwarf galaxy on a
retrograde orbit of low inclination (\citealt{bignone2019}). In terms of
the spatial distribution of accreted stars, stars of a low-mass dwarf
galaxy under the influence of very weak dynamical friction are tidally
stripped off in the outer region of the MW, but its high
eccentricity causes its stars to be deposited in the inner region of the
MW (\citealt{karademir2019}), whereas, due to stronger dynamical
friction, a more-massive dwarf galaxy loses its stars in the inner
region of the MW (\citealt{amorisco2017}).

Following the above reasoning, we can infer that our LOI stars on highly
eccentric and radial orbits may be accreted from two massive dwarf
galaxies with highly eccentric and low-inclination orbits in prograde
and retrograde motions, respectively. Stars stripped from them might
drive the prograde motion in the IHR and retrograde motion in the OHR, due
to different dynamical friction efficacy arising from the different
orbital directions. On the other hand, the HOI stars on highly eccentric
and radial orbits are regarded as tidally stripped ones from one 
massive dwarf galaxy merged on an extremely eccentric and
high-inclination orbit under the influence of dynamical friction.

Meanwhile, the LOI and HOI stars with retrograde motions on highly
eccentric and tangential orbits have the peaks of their MDFs located in
the more metal-poor regime in the OHR than the radial-dominated stars. In
addition, many of those stars have higher $E_{\rm tot}$ than stars on radial
orbits. These aspects lead us to conclude that they may be stripped in
the OHR of the MW, due to weak dynamical friction and self-gravity from
low-mass dwarf galaxies that are on low- and high-inclination orbits
with high eccentricity and retrograde motion at high $E_{\rm tot}$, resulting
in stripped stars that have large \rmax. Taken as a whole, accounting for 
the dynamical characteristics of our LOI and HOI stars require at least
$five$ different accretion episodes.

\begin{figure}
\begin{center}
\epsscale{1.1}
\plotone{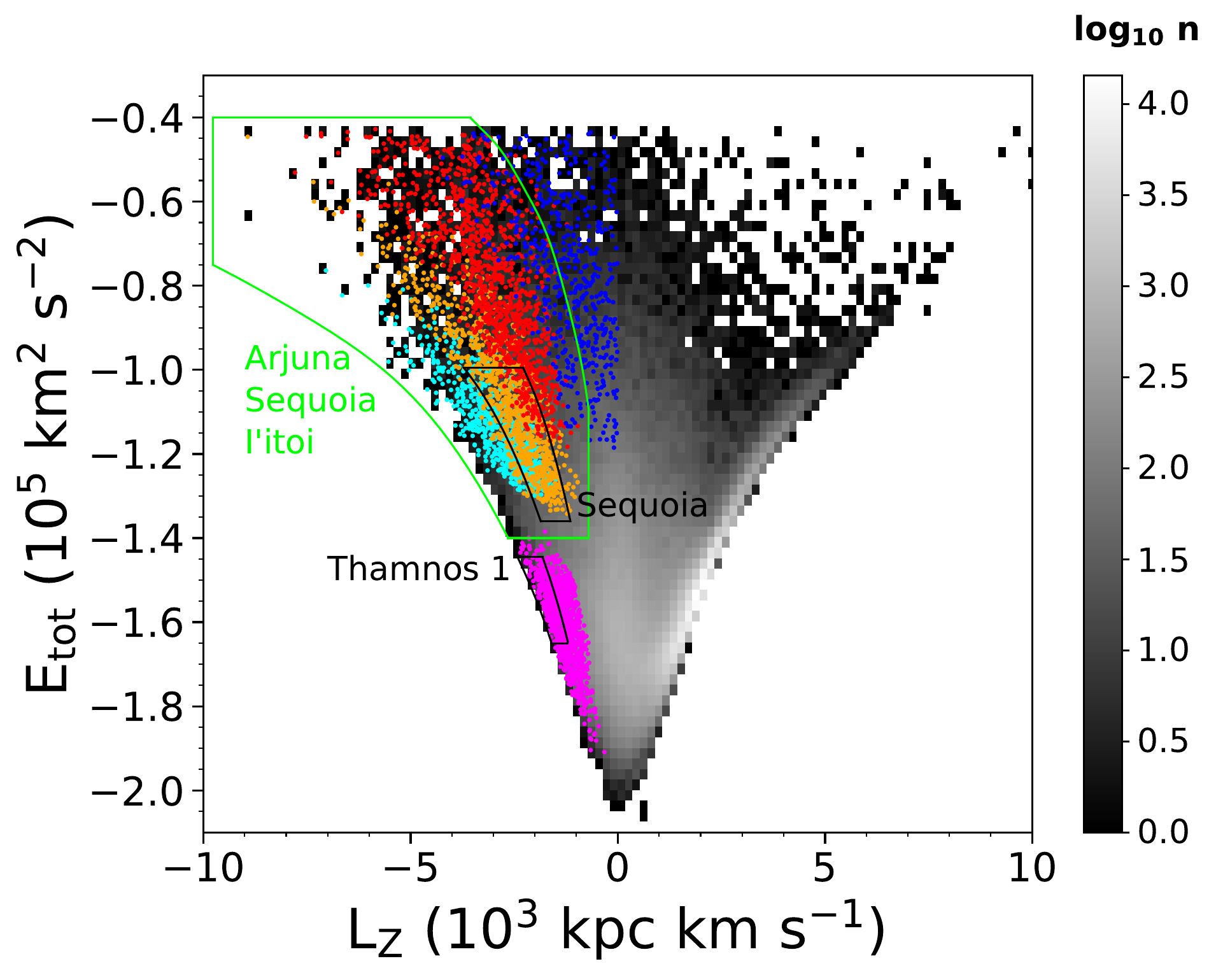}
\caption{Distribution in $E_{\rm tot}$ versus $L_{\rm Z}$ for stars
of previously identified substructures.  Here we considered only stars 
with retrograde motions. Our LOI and HOI stars are displayed in red and blue
colors, respectively. The Sequoia event, with 0.5 $<$ $e$ $\leq$ 0.7, 
$V_{\rm r}^2/V^2 \leq 0.45$, and \rmax\ $\geq$ 20 kpc and Thamnos 1, with $e$ $\leq$ 0.3 
and \rmax\ $<$ 12 kpc, are represented with orange and magenta colors, respectively. The
cyan color marks substructures with $e$ $\leq$ 0.5 and \rmax\ $\geq$ 20
kpc, similar to those reported by \citet{naidu2020}. The green solid box 
indicates the area where the Arjuna, Sequoia, and I'itoi substructures 
noted by \citet{naidu2020} are located. Similarly, black solid boxes mark the 
region of Sequoia and Thamnos 1 by \citet{koppelman2019}.}
\label{figure7}
\end{center}
\end{figure}

To demonstrate that the LOI and HOI stars we analyze are different entities from other
known substructures, we compared our two groups (LOI and HOI) of stars
to those on retrograde orbits identified in literature
(\citealt{koppelman2019,myeong2019,naidu2020}) in the $L_Z$-$E_{\rm{tot}}$ 
diagram shown in Figure~\ref{figure7}. The figure is drawn with the
stars with \rmax\ $<$ 200 kpc, calculated using the \citet{mcmillan2017}
potential for easy comparison with those in literature. Substructures of
stars in the LOI and HOI sub-samples we have identified are represented
in red and blue colors, respectively. The Sequoia event, with 0.5 $<$
$e$ $\leq$ 0.7, $V_{\rm r}^2/V^2 \leq 0.45$, and \rmax\ $\geq$ 20 kpc, 
and Thamnos 1, with $e$ $\leq$ 0.3 and \rmax\ $<$ 12 kpc, are displayed 
with orange and magenta colors, respectively. The cyan color marks sub-samples 
with $e$ $\leq$ 0.5 and \rmax\ $\geq$ 20 kpc, similar to those reported 
by \citet{naidu2020}. From inspection, our LOI and HOI stars on retrograde 
motions are clearly well-separated from the other known substructures.

\section{Summary} \label{sec:summary}

Using metallicities, radial velocities, and distances from SDSS DR12 and
proper motions from $Gaia$ DR2, we have presented an analysis of the
kinematic and orbital properties for MS and MSTO stars with
eccentricities greater than 0.7 and [Fe/H] $<$ --1.0, which are the
typical properties of the GSE stars, after separating them into LOI and 
HOI sub-samples. 

LOI and HOI stars on highly eccentric and radial orbits exhibit different
dynamical characteristics. LOI stars mostly have $e$ $<$ 0.9 and
lower \zmax, whereas HOI stars have $e$ $>$ 0.9 and higher \zmax.
Moreover, LOI stars are separated into two groups with prograde and
retrograde motions, and exhibit prograde motions in the inner halo and
retrograde motions in the outer halo. Accordingly, they are regarded as stars
accreted from two massive dwarf galaxies with prograde and retrograde
orbits of low inclination under the influence of different dynamical
friction due to different orbital directions. On the other hand, HOI stars
globally have a symmetric distribution of rotational velocity near
zero, although they exhibit a small retrograde motion of \vphi\ $\sim$
--15 $\rm{km~s^{-1}}$ in the outer halo. These stars are considered to be stripped
from a massive dwarf galaxy on a orbit of high inclination, based on their
MDF with a peak of [Fe/H] = --1.3.

In addition, our analysis indicates that at least two low-mass
progenitors are required to explain the distinct MDFs and dynamical
properties between the LOI and HOI stars that are on retrograde motion 
with highly eccentric and tangential orbits in the OHR. 

\acknowledgments

We thank an anonymous referee for a careful review of this paper,
which has improved the clarity of its presentation.
Y.S.L. acknowledges support from the National Research Foundation (NRF) of
Korea grant funded by the Ministry of Science and ICT (NRF-2018R1A2B6003961 
and NRF-2021R1A2C1008679). T.C.B. acknowledges partial support for this work
from grant PHY 14-30152; Physics Frontier Center/JINA Center for the Evolution
of the Elements (JINA-CEE), awarded by the U.S. National Science Foundation.

Funding for SDSS-III has been provided by the Alfred P. Sloan Foundation, the
Participating Institutions, the National Science Foundation, and the U.S.
Department of Energy Office of Science. The SDSS-III Web site is
http://www.sdss3.org/.

SDSS-III is managed by the Astrophysical Research Consortium for the Participating 
Institutions of the SDSS-III Collaboration including the University of Arizona, 
the Brazilian Participation Group, Brookhaven National Laboratory, Carnegie Mellon 
University, University of Florida, the French Participation Group, the German 
Participation Group, Harvard University, the Instituto de Astrofisica de Canarias, 
the Michigan State/Notre Dame/JINA Participation Group, Johns Hopkins University,
 Lawrence Berkeley National Laboratory, Max Planck Institute for Astrophysics, 
Max Planck Institute for Extraterrestrial Physics, New Mexico State University,
 New York University, Ohio State University, Pennsylvania State University, 
University of Portsmouth, Princeton University, the Spanish Participation Group, 
University of Tokyo, University of Utah, Vanderbilt University, University of 
Virginia, University of Washington, and Yale University.

\end{document}